\documentstyle{article}

\author{L. S. F. Olavo\\
Departamento de Fisica, Universidade de Brasilia - UnB\\
CEP 70910-900 Brasilia, D.F. - Brazil}
\title{Quantum Mechanics as a Classical Theory\\
IX: The Formation of Operators and\\
Quantum Phase-Space Densities
}

\begin{document}

\maketitle
\begin{abstract}
In our previous papers we were interested in making a reconstruction of
quantum mechanics according to classical mechanics. In this paper we suspend
this program for a while and turn our attention to a theme in the frontier
of quantum mechanics itself---that is, the formation of operators. We then
investigate all the subtleties involved in forming operators from their
classical counterparts. We show, using the formalism of quantum phase-space
distributions, that our formation method, which is equivalent to Weyl's
rule, gives the correct answer. Since this method implies that eigenstates
are not dispersion-free we argue for modifications in the orthodox view.
Many properties of the quantum phase-space distributions are also
investigated and discussed in the realm of our classical approach. We then
strengthen the conclusions of our previous papers that quantum mechanics is
merely an extremely good approximation of classical statistical mechanics
performed upon the configuration space.
\end{abstract}

\section{The Formation of Quantum Operators}

The formation problem of quantum mechanical operators has already been
treated by a number of authors \cite{1,2,3,4,5,6}. The problem resides
basically in forming, from a given classical function $f(q,p)$ of the
commuting generalized coordinates and momenta, the related quantum
mechanical operator $O[f(p,q)]$.

This problem, as expressed in Shewell's review article \cite{3}, either
cannot be solved in a non-ambiguous way (e.g. von Neumann's and Dirac's
rules) or it does not give the result expected by the orthodox epistemology
of quantum mechanics, viz. Weyl's and Revier's rules.

As an example, the operators related to the classical function $f(q,p)=q^2p^2
$ are
\begin{equation}
\label{1}\mbox{von Neumann: }\left\{
\begin{array}{c}
O[q^2p^2]=O[q]^2O[p]^2-2i\hbar O[q]O[p]-\hbar ^2/4 \\
O[q^2p^2]=O[q]^2O[p]^2-2i\hbar O[q]O[p]-\hbar ^2
\end{array}
\right.
\end{equation}
and
\begin{equation}
\label{2}\mbox{Dirac: }\left\{
\begin{array}{c}
O[q^2p^2]=O[q]^2O[p]^2-2i\hbar O[q]O[p]-\hbar ^2/3 \\
O[q^2p^2]=O[q]^2O[p]^2-2i\hbar O[q]O[p]-2\hbar ^2/3
\end{array}
\right.
\end{equation}
which are ambiguous and so, contrary to the postulates of quantum
mechanics---each `observable' must be assigned to one and only one operator.

For Weyl's rule we have
\begin{equation}
\label{3}O[q^2p^2]=O[q]^2O[p]^2-2i\hbar O[q]O[p]-\hbar ^2/2
\end{equation}
which is unambiguous but has the drawback that, for the harmonic oscillator
problem with the hamiltonian given by
\begin{equation}
\label{4}H=\frac 12(p^2+q^2),
\end{equation}
this rule gives (figure I)
\begin{equation}
\label{5}O[H^2]=\{O[H]\}^2+\hbar ^2/4
\end{equation}
which is equivalent to saying that it predicts energy dispersions for the
energy eigenstates (that is also contrary to the orthodox epistemological
considerations of quantum mechanics). In this example, equation (\ref{3})
says that the dispersion will be given by
\begin{equation}
\label{6}\left( \Delta E\right) ^2=<O[H^2]-<O[H]>>^2=\hbar ^2/4,
\end{equation}
for all energy levels.

Weyl's rule has also the disadvantage that some functions of classical
constants of motion are transformed into operators which do not commute with
the hamiltonian and are not constants of motion in the quantum mechanical
side. This may be shown using the hamiltonian\cite{3}
\begin{equation}
\label{7}H=p^2/2+q^4/4,
\end{equation}
for which, Weyl's rule gives
\begin{equation}
\label{8}O[H^2]=\{O[H]\}^2+3\hbar ^2q^2/4,
\end{equation}
implying that $O[H^2]$ does not represent a constant of motion, since it
does not commute with $H$.

In short, the acceptance of Weyl's rule implies that we do not accept the
first von Neumann's rule \cite{3}
\begin{equation}
\label{9}\mbox{if }O[A]={\bf A}\Rightarrow O[f(A)]=f({\bf A}).
\end{equation}

It is possible to show that the symmetrization rules may be reduced to
Weyl's rule and present the same problems.

It is the aim of the next section to show, using our formalism\cite
{7,8,9,10,11,12,13,14} why ambiguities arrive and to throw some confidence
upon Weyl's rule, which is equivalent to our own.

In the third section, we show that this result is in accordance with the
results obtained using the classical phase-space functions derived using our
previously defined formalism \cite{7,8,9,10,11,12,13,14}.

Some of the main characteristics of these functions will be dealt with in
the fourth section by means of some illustrative examples that will throw
some light upon the discussion.

\section{The Ambiguity Problem}

It has been already shown that the ambiguity related to von Neumann's and
Dirac's rules (to cite but a few) are related with the attempt of mapping a
commutative ring into a non-commutative one \cite{2}. We will show this
using our analytical method.

{}From our previous work \cite{7} one might see that the Schr\"odinger
equation of quantum mechanics is derived from the classical Liouville
equation using the Infinitesimal Wigner-Moyal Transformation
\begin{equation}
\label{11}\rho \left( q-\frac{\delta q}2,q+\frac{\delta q}2;t\right) =\int
F(q,p;t)e^{\frac i\hbar p\delta q}dp,
\end{equation}
where $F(q,p;t)$ is the classical phase-space probability density function
and $\rho \left( q,\delta q;t\right) $ is the quantum density function (also
called density matrix) and where $\delta q$ is an infinitesimal increment
{\it taken over the coordinate q-axis}.

To derive Schr\"ondiger equation it is necessary to impose upon the density
function the functional form
\begin{equation}
\label{12}\rho \left( q-\frac{\delta q}2,q+\frac{\delta q}2;t\right) =\psi
^{\dagger }\left( q-\frac{\delta q}2;t\right) \psi \left( q+\frac{\delta q}%
2;t\right)
\end{equation}
and to expand this product with
\begin{equation}
\label{13}\psi \left( q;t\right) =R(q;t)e^{iS(q;t)/\hbar }
\end{equation}
in powers of the infinitesimal increment $\delta q$. We then take the
density (\ref{12}) into its equation and, since this increment is
infinitesimal, we keep only the zeroth and first order resulting terms and
show that the former implies the continuity equation while the later implies
Schr\"odinger equation.

It is noteworthy that in this procedure we have lost one of the independent
variables, viz. the infinitesimal $\delta q$ related to $p$. This makes our
problem much simpler but it has the disadvantage of introducing the
non-commutativity of the canonically conjugated quantities. The explanation
why this formalism is rather triumphant is related with the fact that the
error introduced with non-commutativity is of the order of Planck's constant
and so, the method is expected to be efficient to within this approximation
(which is rather good).

The construction of operators is a direct result of the definition (\ref{11}%
). Since the mean values of a classical function $f(q,p)$ are given by
\begin{equation}
\label{14}<f(q,p)>=\int \int f(q,p)F(q,p;t)dqdp,
\end{equation}
it is automatic that the mean of the operator related with this function is
given by\cite{7}
\begin{equation}
\label{15}<O[f(q,p)]>=\int \lim _{\delta q\rightarrow 0}f\left( \widehat{q}%
,-i\hbar \frac \partial {\partial (\delta q)}\right) \rho \left( q-\frac{%
\delta q}2,q+\frac{\delta q}2;t\right) dx
\end{equation}
giving the same values of expression (\ref{14}).

The function $f\left( \widehat{q},-i\hbar \frac \partial {\partial (\delta
q)}\right) $ is unambiguous defined since $q$ and $\delta q$ are independent
variables.

Now, giving the amplitudes $\psi \left( q;t\right) $, we can form $\rho
\left( q-\frac{\delta q}2,q+\frac{\delta q}2;t\right) $ according to
expression (\ref{12}) and expand these amplitudes in a Taylor series around $%
q$ to get, until second order in $\delta q$
$$
<O[f(q,p)]>=\int \lim _{\delta q\rightarrow 0}f\left( \widehat{q},-i\hbar
\frac \partial {\partial (\delta q)}\right) \left[ \psi ^{\dagger }\psi
+\left( \frac{\delta q}2\right) \left( \psi ^{\dagger }\frac{\partial \psi }{%
\partial q}-\frac{\partial \psi ^{\dagger }}{\partial q}\psi \right)
+\right.
$$
\begin{equation}
\label{16}\left. +\frac 1{2!}\left( \frac{\delta q}2\right) ^2\left( \psi
^{\dagger }\frac{\partial ^2\psi }{\partial q^2}+\frac{\partial ^2\psi
^{\dagger }}{\partial q^2}\psi -2\frac{\partial \psi }{\partial q}\frac{%
\partial \psi ^{\dagger }}{\partial q}\right) +o(\delta q^3)\right] ,
\end{equation}
and then apply to each term of the expansion the operator $\widehat{f}$
defined in (\ref{15}) and take the limit as indicated.

The procedure defined above is unambiguous and, it might be easily shown,
gives the same results of Weyl's rule. We have, for example, in the case of $%
f(q,p)=q^2p^2$, the operator
\begin{equation}
\label{17}\widehat{f}\left( \widehat{q},-i\hbar \frac \partial {\partial
(\delta q)}\right) =\lim _{\delta q\rightarrow 0}\widehat{q}^2\left( -\hbar
^2\frac{\partial ^2}{\partial (\delta q)^2}\right)
\end{equation}
giving, in terms of the amplitudes, the operator
\begin{equation}
\label{18}\widehat{f}=\widehat{q}^2\widehat{p}^2-2i\hbar \widehat{q}\widehat{%
p}-\hbar ^2/2.
\end{equation}

However, in expression (\ref{15}), we could have considered the term $q^2$
as part of the function $\rho \left( q-\frac{\delta q}2,q+\frac{\delta q}%
2;t\right) $; in the example above this means that we are calculating
\begin{equation}
\label{19}<O[f(q,p)]>=\int \lim _{\delta q\rightarrow 0}\left( -\hbar ^2
\frac{\partial ^2}{\partial (\delta q)^2}\right) \left[ q^2\rho \left( q-
\frac{\delta q}2,q+\frac{\delta q}2;t\right) \right] dq,
\end{equation}
which is allowed since $q$ and $\delta q$ are independent variables. In this
case we will expand the function
\begin{equation}
\label{20}q^2\rho \left( q-\frac{\delta q}2,q+\frac{\delta q}2;t\right)
=q^2\psi ^{\dagger }\left( q-\frac{\delta q}2;t\right) \psi \left( q+\frac{%
\delta q}2;t\right)
\end{equation}
and there are three possibilities of doing that
\begin{equation}
\label{21}\left\{
\begin{array}{l}
\{q^2\psi ^{\dagger }\left( q-
\frac{\delta q}2;t\right) \}\psi \left( q+\frac{\delta q}2;t\right)
\longmapsto \widehat{q}^2\widehat{p}^2-2i\hbar \widehat{q}\widehat{p}-\hbar
^2 \\ \psi ^{\dagger }\left( q-
\frac{\delta q}2;t\right) \{q^2\psi \left( q-\frac{\delta q}2;t\right)
\}\longmapsto \widehat{q}^2\widehat{p}^2-2i\hbar \widehat{q}\widehat{p}%
-\hbar ^2 \\ \{q\psi ^{\dagger }\left( q-\frac{\delta q}2;t\right) \}\{q\psi
\left( q-\frac{\delta q}2;t\right) \}\longmapsto \widehat{q}^2\widehat{p}%
^2-2i\hbar \widehat{q}\widehat{p}
\end{array}
\right.
\end{equation}
or we could make the expansion of
\begin{equation}
\label{22}q\left( -\hbar ^2\frac{\partial ^2}{\partial (\delta q)^2}\right)
\left[ q\psi ^{\dagger }\left( q-\frac{\delta q}2;t\right) \psi \left( q+
\frac{\delta q}2;t\right) \right] \}
\end{equation}
giving two other possibilities.

All these results are different from the previous one (\ref{18}) because the
infinitesimal increment is taken over the coordinate $q$-axis and taking the
limit $\delta q\rightarrow 0$ implies making one variable to disappear.

Another way to see that is making the coordinate transformation
\begin{equation}
\label{23}y=q+\frac{\delta q}2\mbox{ ; }y^{\prime }=q-\frac{\delta q}2
\end{equation}
giving
\begin{equation}
\label{24}\frac \partial {\partial (\delta q)}=\left( \frac \partial
{\partial y}-\frac \partial {\partial y^{\prime }}\right) \ ;\ \rho =\rho
(y^{\prime },y;t)
\end{equation}

We then have
\begin{equation}
\label{25}\rho (y^{\prime },y;t)=\sum_{n=0}^\infty \frac 1{n!}\left[
(y-y_0)\frac \partial {\partial y}+(y^{\prime }-y_0^{\prime })\frac \partial
{\partial y^{\prime }}\right] ^n\rho (y_0^{\prime },y_0;t)
\end{equation}
where it is implicit that we first perform the derivatives and then compute
their values at
\begin{equation}
\label{25a}y_0=y_0^{\prime }=q
\end{equation}
Using expression (\ref{24}) we get for the mean value of the classical
phase-space monomial represented by $q^Np^M$ the corresponding expression%
$$
<q^Np^M>=(-i\hbar )^M\int \frac{(y+y^{\prime })^N}{2^N}\left[ \frac \partial
{\partial y}-\frac \partial {\partial y^{\prime }}\right] ^M\cdot
$$
\begin{equation}
\label{27}\cdot \sum_{n=0}^\infty \frac 1{n!}\left[ (y-y_0)\frac \partial
{\partial y}+(y^{\prime }-y_0^{\prime })\frac \partial {\partial y^{\prime
}}\right] ^n\rho (y_0^{\prime },y_0;t)d\left[ \frac{y+y^{\prime }}2\right] .
\end{equation}

Now, since $y^{\prime }$ and $y$ are independent coordinates we might change
the order of some of the terms in the above expression which are of the form
\begin{equation}
\label{28}y^k\frac \partial {\partial y^{\prime j}}\ ;\ y^{\prime k}\frac
\partial {\partial y^j}
\end{equation}
but when we calculate the limit we see that, in this limit,
\begin{equation}
\label{29}y=y^{\prime }=q
\end{equation}
and are not independent quantities anymore (they are, indeed, equal). This
limiting process is exactly the one we have used, as described above, to
derive the Schr\"odinger equation and is also, as already shown\cite{7} the
one responsible for the introduction of non-commutativity of the dynamic
operators.

We are not able to choose, at this place, one or another of the procedures
as shown in the examples (\ref{20},\ref{21}) and (\ref{22}) but the claims
for the procedure leading to Weyl's rule are obvious. This is because it is
not clear why we will expand that part of the operator (we are trying to
find an expression for) as part of the density function; indeed, in
expression (\ref{17}), for example, the term $q^2$ refers to an {\it operator%
} while the density function is indexed by $q$, a {\it coordinate} variable.

We must be aware however, of the consequences of this choice. One of them,
for example, will be to predict, for the harmonic oscillator example, a
dispersion in the energy according to expression (\ref{5}) of
\begin{equation}
\label{29a}\Delta E=\frac \hbar 2,
\end{equation}
a result to which we will soon return.

In the next section, we will show that these claims for Weyl's rule is
strengthened when we return, by means of expression (\ref{11}), to the
classical phase-space formalism. We will also discuss some of the properties
of these functions in the realm of the present developments.

\section{Classical Phase-Space Distributions}

Distributions of probability defined upon classical phase-space were first
proposed by Wigner\cite{15} in connection with thermodynamics
considerations. Latter, the theme was further developed by Moyal\cite{16}
and Groenewold\cite{2}. In the late sixties, the problem began again to
deserve some attention mainly in connection with problems on the coherent
properties of the electromagnetic field\cite{5}, on the theory of laser and
on non-linear processes\cite{1,6}.

More recently, we have seen the revival of the problem of writing quantum
mechanics using these distributions and the notion of stochastic phase-space
\cite{p1,p2,p3,p4,p5,p6,p7,p8,p9,p10,p11}.

All these latter developments are rather technical and pay more attention to
formal aspects of these distributions than to the epistemological
considerations that could follow from these developments. Such
considerations were indeed well formulated in Moyal's paper where the
epistemological deficiencies related to the tentative of linking classical
and quantum mechanics by writing the later in classical phase-space were
pointed out.

The major problem with the distributions $F(q,p;t)$, derived from
expressions nearly similar to (\ref{11}) and defined over classical
phase-space---as recognized by all the cited authors---, is the fact that
these functions might take negative values; they are then called
quasi-probability functions. This fact is usually interpreted as meaning our
necessarily introduction of an error when passing from a fundamental theory
(quantum mechanics with non-commuting `observables') to an approximate
theory (classical statistical mechanics with commuting dynamic variables).

In the context of the present series of papers this conclusion has to be
reversed. We have already shown that it is possible to derive quantum
mechanics from classical mechanics without introducing any non-classical
postulate. We have also shown that this process of derivation, where one of
the phase-space coordinates (usually the momentum) disappears, introduces
the non-commutativity when we impose that our densities have to obey
equation (\ref{12}). The hierarchy of the theories is fixed by this
derivation and we see that quantum mechanics will give good results whenever
the experimental environment is not able to uncover Nature with techniques
where the dispersion product is given by
\begin{equation}
\label{30}\Delta q\Delta p<\frac \hbar 2,
\end{equation}
where $q$ and $p$ are the generalized canonically conjugate coordinate and
momentum. That is to say: {\it the dispersion relations do not impose any
constraint upon the behavior of Nature but only upon our capacity of
describing Nature by means of quantum theory}.

Because of the smallness of Planck's constant, we might conclude that this
technique is rather fruitful and the error introduced by the quantum
formalism is far from being detectable with nowadays laboratory technology.

Then, as we have already shown, the passage from a description based upon
the classical density function to a description based upon the density
`matrix' $\rho (q,\delta q)$ is exact but we still have two independent
variables and, as in the classical phase-space, no dispersions\cite{7}. When
passing from densities $\rho (q,\delta q)$ to a description using the
amplitudes $\psi (q;t)$, the dispersion relation becomes
\begin{equation}
\label{31}\Delta q\Delta p\geq \frac \hbar 2,
\end{equation}
but the problem becomes rather tractable and, as the error introduced is
absurdly small, we might conclude that more was gained than was lost.

Now, when trying to return to classical phase-space---not by inverting the
Fourier Transform (\ref{11}) but by finding the function $F(q,p;t)$ that,
when integrated will result in the desired $\rho (q,\delta q)$\cite{9}---we
shall admit that the introduced error related with the dispersion relation
and with the imposition (\ref{12}) will propagate upward and will
necessarily be incorporated into our densities $F(q,p;t)$. This fact might
be easily verified from the usual expression\cite{16} for the classical
density
\begin{equation}
\label{32}F(q,p;t)=e^{-\frac 12i\hbar \partial ^2/\partial p\partial
q}\left[ \psi ^{\dagger }(q)\phi (p)e^{ipq/\hbar }\right] ,
\end{equation}
where $\psi (q)$ and $\phi (p)$ are the position and momentum
eigenfunctions. This expression does indeed incorporate\cite{16}
Heisemberg's inequality.

The fact that such functions are not positive definite might be taken as an
indication that the process of quantization---equation (\ref{12}) and the
limit $\delta x\rightarrow 0$--- indeed introduces an error.

Another problem with these distributions is the fact that, although they
give the correct values for the energies mean values and the dispersion
relations for $q$ and $p$, they also predict a dispersion in the energy
eigenstates---something rather contrary to the orthodox interpretation where
eigenstates are dispersionless.

This is the point we wanted to reach that will help us with the discussions
about which method of operator construction to choose.

In the next section we will develop some examples of classical phase-space
distributions for quantum systems and will study more deeply the harmonic
oscillator problem to stress some relevant points for our latter discussion.

\section{Some Examples: The Harmonic Oscillator}

We begin by giving some obvious examples:

(i) Free Particle:
\begin{equation}
\label{33}\psi (x,t)=e^{ikx}\Rightarrow \rho (x,\delta x)=e^{ik\delta
x}\Rightarrow F(x,p;t)=\delta _D(k-p/\hbar )
\end{equation}
where $\delta _D(x)$ is the Dirac's delta function. This density $\rho $$%
(x,\delta x)$ shall be compared to the one obtained when passing from the
classical phase-space to the `quantum' $(x,\delta x)$ configuration space.
Indeed, we have already shown\cite{9} that, for a dispersion free {\it %
ensemble} of systems with one free particle, where the classical density
function is given by
\begin{equation}
\label{33.1}F(x,p;t)=\delta _D[x-x_0(t)]\delta _D[p-p_0(t)],
\end{equation}
the correct density function is given by
\begin{equation}
\label{33.a}\rho (x,\delta x)=\delta _D[x-x_0(t)]e^{ik\delta x}
\end{equation}
and represents an {\it ensemble} of systems with one localized free
particle. Since this density function represents a dispersion free {\it %
ensemble} it cannot be expressed within quantum theory based upon
Schr\"odinger's equation. We thus clearly see that, when passing from the
quantum to the classical, this localization cannot be retrieved and we are
left with expression (\ref{33}) which has no information about the position
of the particle---this information was lost.

(ii) Minimum Free Wave-Packet:
\begin{equation}
\label{34}\psi (x,t)=e^{-x^2/2\alpha ^2+ik_ox}\Rightarrow \rho (x,\delta
x)=e^{-x^2/2\alpha ^2}e^{-(\delta x)^2/4\alpha ^2+ik_0\delta x},
\end{equation}
where $a=\Delta x\sqrt{2}$. We then have
\begin{equation}
\label{35}F(x,p;t)=Ne^{-\left[ x^2/2\alpha ^2+(p-\hbar k_0)^2a^2/\hbar
^2\right] },
\end{equation}
which implies that $\Delta p=\hbar \sqrt{2}/a$ and so
\begin{equation}
\label{36}\Delta x\Delta p=\hbar /2.
\end{equation}

We now want to develop another example more deeply.

(iii)The Harmonic Oscillator: The probability amplitudes are given by
\begin{equation}
\label{37}\psi _n(x)=\left[ \frac 1{2^nn!\pi ^{1/2}}\right]
^{1/2}H_n(x)e^{-\frac 12x^2},
\end{equation}
where $H_n(x)$ is the n-th Hermite polynomial. These amplitudes are solution
of the Schr\"odinger equation
\begin{equation}
\label{38}\widehat{E}\psi _n(x)=\frac 12\left[ \widehat{p}^2+\widehat{q}%
^2\right] \psi _n(x)=E_n\psi _n(x)=(n+1/2)\hbar \psi _n(x),
\end{equation}
where $\widehat{E}$ is the hamiltonian operator and $E_n$ its eigenvalues.

Using expression (\ref{11}) we might explicitly calculate $F(x,p;t)$ to get
the normalized functions
\begin{equation}
\label{39}\left\{
\begin{array}{l}
n=0\qquad F_0(x,p;t)=\frac 1{\pi \hbar }e^{-H} \\
n=1\qquad F_1(x,p;t)=\frac 2{\pi \hbar }\left[ H-\frac 12\right] e^{-H} \\
n=2\qquad F_2(x,p;t)=\frac 2{\pi \hbar }\left[ \left( H-1\right) ^2-\frac
12\right] e^{-H} \\
n=3\qquad F_3(x,p;t)=\frac 2{3\pi \hbar }\left[ 2\left( H-2\right)
^3+(H-2)^2-3(H-2)-\frac 72\right] e^{-H}
\end{array}
\right. ,
\end{equation}
where
\begin{equation}
\label{41}H=x^2+p^2/\hbar ^2.
\end{equation}

In figures II,III and IV we plot the choices $n=0,3$ and $10$ respectively.
It is clear that, apart from the ground state, we have the appearing of
negative probability densities.

As amply noticed in the literature, we get non-negative densities when we
integrate in one or other phase-space coordinates giving the lateral
probabilities
\begin{equation}
\label{42}F_1(x;t),F_2(p;t)\geq 0\ \forall x\mbox{ and }\forall p
\end{equation}
which are all in accordance with the quantum mechanical results.

With the density functions $F(x,p;t)$ at hand we might calculate the mean
energies, the Heisemberg dispersion relations and the mean quadratic
deviation (the dispersion) of these energies. We calculated these values for
the first $n=0,..,7$ functions and quoted the results in table I.

Looking at this table we can see that the energy mean values and the
Heisemberg dispersion relations are given by
\begin{equation}
\label{43}<E_n>=(n+1/2)\hbar \mbox{ and }\Delta x\Delta p=(n+1/2)\hbar
\end{equation}
as expected\cite{Schiff}. For the energy mean quadratic deviation\ we get
\begin{equation}
\label{44}\Delta E=\frac \hbar 2
\end{equation}
for all states.

We now return to expression (\ref{4}) and see that result (\ref{44}) agrees
with the way we define operators using our formalism (and also with Weyl's
rules). It is important to stress that, when finding $F(x,p;t)$ {\it from}
the quantum mechanical amplitudes we have not made any approximations---we
have just performed the `inversion' indicated by (\ref{11}). The value of
the energy dispersion shall agree with the quantum mechanical one as in the
cases of the mean energy and Heisemberg's relations (\ref{43}).

We then must abandon the first von Neumann's rule\cite{3} represented in
expression (\ref{9}) and also conclude that even eigenstates of the
Schr\"odinger equation are not dispersion-free.

The results above are good indications that our operator construction
procedure, and also Weyl's rule, is the correct one to be used in the formal
structure of quantum mechanics.

Another consequence of expression (\ref{44}), which may be written generally
as
\begin{equation}
\label{44.1}\Delta E=\frac \hbar 2\omega ,
\end{equation}
where $\omega $ is the frequency of the harmonic oscilator (we have made
equal to unity in the previous development), is the conclusion, based on the
Heisemberg's energy-time dispersion relation, that the dispersion in time
must be given by
\begin{equation}
\label{44.2}\Delta t=\frac 1\omega =\frac T{2\pi },
\end{equation}
where $T$ is the period of oscilation.

We also note that, considering expressions (\ref{43}) and (\ref{44.1}), the
correct energies (within the mean quadratic deviations) of the harmonic
oscilator problem shall be written as
\begin{equation}
\label{44.3}E_n=<E_n>\pm \Delta E=(n+\frac 12)\hbar \mbox{ }\pm \frac \hbar
2,
\end{equation}
which throws some light upon the meaning of the vacuum energy fluctuation
(we will return to this problem in a latter work).

\section{Constants of Motion}

The acceptance of our rules of operator formation leads to the problem
eluded in (\ref{7}) where functions of constants of motion are not
necessarily constants of motion---since, now, some function of the
hamiltonian operator, say $f(H)$, might not even commute with $H$.

We might see this in a straightforward way if we look for the equations
satisfied by the density functions $F(q,p;t)$. If we write $F^Q(q,p;t)$ for
the function obtained from the quantum mechanical amplitudes, as above, and
write $F^C(q,p;t)$ for the classical distribution, we have, by hypothesis
\begin{equation}
\label{42a}\frac{\partial F^C}{\partial t}+\{H,F^C\}=0
\end{equation}
while, it is possible to show\cite{16}, the equation obeyed by function $%
F^Q(q,p;t)$ is given by
\begin{equation}
\label{42b}\frac{\partial F^Q}{\partial t}+\frac 2\hbar \sin \frac \hbar
2\{\frac \partial {\partial p},\frac \partial {\partial
q}\}H(q,p)F^Q(q,p;t)=0.
\end{equation}

This makes clear our point of view. We began with equation (\ref{42a}) and
end with another different equation (\ref{42b}). This is another argument
for concluding that we shall not have
\begin{equation}
\label{43a}F^C(q,p;t)=F^Q(q,p;t).
\end{equation}

In some special cases (the harmonic oscillator is an example) equation (\ref
{42a}) is equivalent to equation (\ref{42b}). Even in such cases we shall
not have (\ref{43a}) since we have postulated, in passing from the former to
the later, one specific functional form for our densities and also a
dispersion relation given by (\ref{31}).

It is also important to stress that the `classical limit' cannot be
generally attained by taking the limit $\hbar \rightarrow 0$ as usually
stated in the literature. This was also shown elsewhere\cite{Cohn}.

The important result (\ref{42b}) means exactly what we have said above.
Indeed, even if some quantity does commute with the Hamiltonian operator,
some functional of it (e.g. its powers) will not generally satisfy this
requirement and will not be a constant of motion\cite{16}. This is another
result that indicates that expression (\ref{42b}) must be only an
approximation. This was already exemplified above in expressions (\ref{7})
and (\ref{8}), where we have noted that the square of the hamiltonian does
not commute with the hamiltonian itself and cannot be a constant of motion.

This same result might be exemplified from Dirac's rule. Since this rule
uses precisely the substitution of the classical Poisson brackets by the
commutator\cite{3} and does not produce, in general, unambiguous operators,
the one-to-one correspondence of classical and quantum-mechanical constants
of motion generally fails.

\section{Conclusions}

Until now we have been concerned with reconstructing quantum mechanics from
classical mechanics\cite{7,8,9,10,11,12,13,14}. This task was taken from the
mathematical and epistemological points of view and there are, still, much
work awaiting---for example, the reconstruction of quantum field theory.

This paper, however, has undertaken a different approach. The theme we have
chosen to approach here lies in the frontier of applicability of the quantum
formalism itself.

Indeed, one of the most fundamental axioms of the orthodox {\it %
interpretation} of quantum theory is the one-to-one correspondence between
classical functions (to be quantized) and quantum operators (observables).
We have thus avoided to contradict this postulate since it comes naturally
from our own derivation of quantum mechanics.

We were then led to adopt Weyl's procedure, or the one we have derived from
our own theory, as the correct one for the formation of quantum mechanical
operators. The most striking consequence of our choice was to be left with
energy eigenstates presenting energy dispersions.

We then have shown that the appearance of these dispersions are pertinent to
the formalism since we derived them from the quantum phase-space
distributions $F^Q(x,p;t)$, defined above, which keep the same physical
content (they are only Fourier transforms of the quantum density function)
of the quantum distributions.

The unavoidable appearance of energy eigenstates with energy dispersion
points in the direction of an interpretation of these states very different
from the one given by the orthodox epistemological approach. We must
abandon, for example, the idea that when performing a measurement upon an
energy eigenstate we do not disturb this energy value and the next
measurement will observe the same value for this variable.

Concerning the formalism, the adoption of our procedure of operator
formation leads to the physically unacceptable result according to which
functions of constants of motion are not themselves constants of motion. It
is interesting to note, however, that this pathological behavior comes from
the attempt to map a commutative ring into a non-commutative one, and is of
the order of, at least, the square of Planck's constant; since the magnitude
of this error is highly beyond experimental precision, we may still expect
this pathology not to be easily empirically detectable---this, of course,
does not preclude us from studying the theme, since we are interested here
in placing the quantum {\it theory} on the most solid ground. All these
results might be also derived from the quasi-Liouville equation (\ref{42b})
obeyed by the quantum phase-space distributions above.

\newpage

\begin{table}
 \begin{center}
  \begin{tabular}{|c|c|c|c|} \hline
     {\bf n}  &  {\bf energy}  &  $\Delta q\Delta p$ & $ \Delta E $ \\ \hline
         0      &     1/2           &               1/2          &       1/2
    \\
         1      &     3/2           &               3/2          &       1/2
    \\
         2      &     5/2           &               5/2          &       1/2
    \\
         3      &     7/2           &               7/2          &       1/2
    \\
         4      &     9/2           &               9/2          &       1/2
    \\
         5      &    11/2          &               11/2         &       1/2
   \\
         6      &    13/2           &               13/2        &       1/2
   \\
         7      &    15/2           &               15/2        &       1/2
   \\ \hline
  \end{tabular}
  \caption{Mean energies and dispersions for the harmonic oscilator derived
                from the quantum phase-space distributions.}
 \end{center}
\end{table}

\newpage

\unitlength=1.00mm \special{em:linewidth 1pt} \linethickness{1pt}

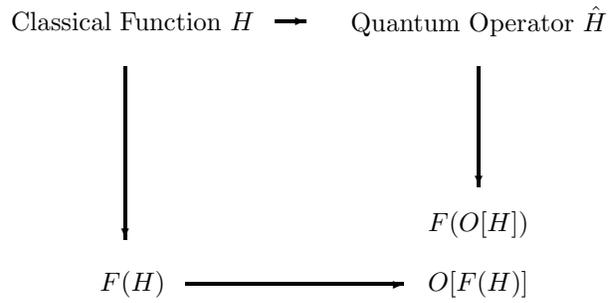
\begin{figure}[p]
\begin{center}
   \begin{picture}(75.00,50.00)
   \put(29.00,50.00){\makebox(0,0)[cc]{Classical Function $H$}}
   \put(75.00,50.00){\makebox(0,0)[cc]{Quantum Operator $\hat{H}$}}
   \put(75.00,44.00){\vector(0,-1){16.00}}
   \put(75.00,23.00){\makebox(0,0)[cc]{$F(O[H])$}}
   \put(28.00,44.00){\vector(0,-1){23.00}}
   \put(29.00,15.00){\makebox(0,0)[cc]{$F(H)$}}
   \put(75.00,15.00){\makebox(0,0)[cc]{$O[F(H)]$}}
   \put(48.00,50.00){\vector(1,0){4.00}}
   \put(36.00,15.00){\vector(1,0){29.00}}
   \end{picture}
\caption{Graph showing the fail of von Neumann's process of quantization.}
\end{center}
\end{figure}


\begin{thebibliography}{99}
\bibitem{1}  Agarwal, G. S. and Wolf, E., Phys. Rev. D {\bf 2},
2161,2187,2206 (1970).

\bibitem{2}  Groenewold, H. J., Physica {\bf 12}, 405 (1946).

\bibitem{3}  Shewell, J. R., Am. J. Phys. {\bf 27}, 16 (1959).

\bibitem{4}  Pool, J. T., J. Math. Phys. {\bf 7}, 66 (1966).

\bibitem{5}  Metha, C. L. and Sudarshan, E. C. G., Phys. Rev. {\bf 138},
B274 (1965).

\bibitem{6}  Misra, S. P. and Shankara, T. S., J. Math. Phys. {\bf 9}, 299
(1968).

\bibitem{7}  Olavo, L. S. F., quant-ph/9503020

\bibitem{8}  Olavo, L. S. F., quant-ph/9503021

\bibitem{9}  Olavo, L. S. F., quant-ph/9503022

\bibitem{10}  Olavo, L. S. F., quant-ph/9503024

\bibitem{11}  Olavo, L. S. F., quant-ph/9503025

\bibitem{12}  Olavo, L. S. F., quant-ph/9509012

\bibitem{13}  Olavo, L. S. F., quant-ph/9509013

\bibitem{14}  Olavo, L. S. F., quant-ph/9511028

\bibitem{15}  Wigner, E. P., Phys. Rev. {\bf 40}, 749 (1932).

\bibitem{16}  Moyal, J. E., Proc. Cambridge Phil. Soc. {\bf 45}, 99 (1949).

\bibitem{Schiff}  Schiff, L., ''{\it Quantum Mechanics}'', (McGraw-Hill Book
Co., Singapore, 1965).

\bibitem{Cohn}  Cohn, J., Am. J. Phys. {\bf 40}, 463 (1972).

\bibitem{p1}  Ali, S. and Prugovecki, E., J. Math. Phys. {\bf 18}, 219
(1977).

\bibitem{p2}  Ali, S. and Prugovecki, E., Physica {\bf 89A}, 501 (1977).

\bibitem{p3}  Ali, S. and Prugovecki, E., Intern. J. of Theor. Phys. {\bf 16}%
, 689 (1977).

\bibitem{p4}  Prugovecki, E., Ann. of Phys. {\bf 110}, 102 (1978).

\bibitem{p5}  Prugovecki, E., Physica {\bf 91A}, 202 (1978).

\bibitem{p6}  Prugovecki, E., Physica {\bf 91A}, 229 (1978).

\bibitem{p7}  Prugovecki, E., J. Math. Phys. {\bf 19}, 2260 (1978).

\bibitem{p8}  Prugovecki, E., J. Math. Phys. {\bf 19}, 2271 (1978).

\bibitem{p9}  Prugovecki, E., Phys. Rev. D {\bf 18}, 3655 (1978).

\bibitem{p10}  Prugovecki, E., Found. of Phys. {\bf 9}, 575 (1979).

\bibitem{p11}  Ali, S., Gagnon, R. and Prugovecki, E., Can. J. Phys. {\bf 59}%
, 807 (1981).
\end{thebibliography}
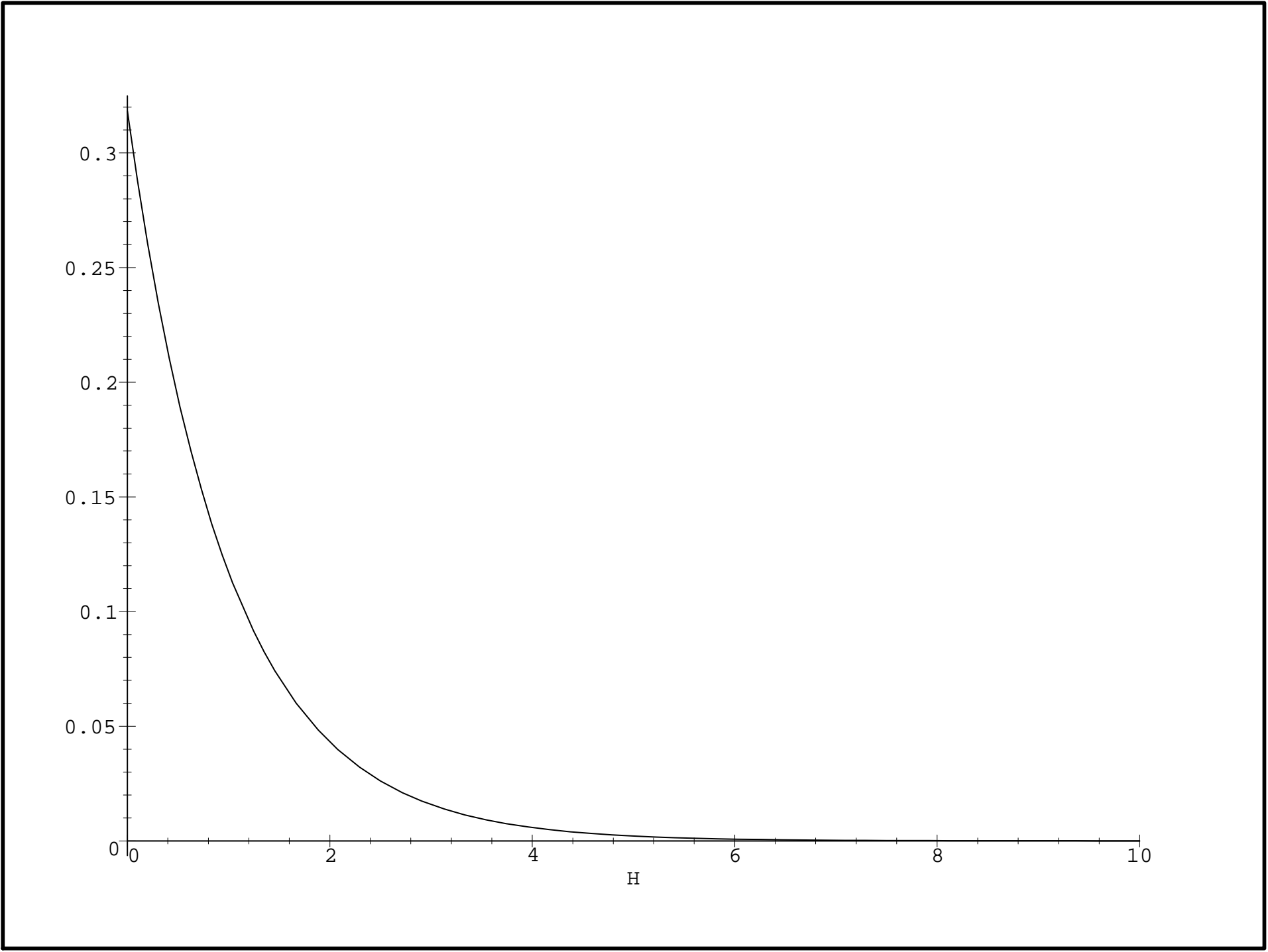
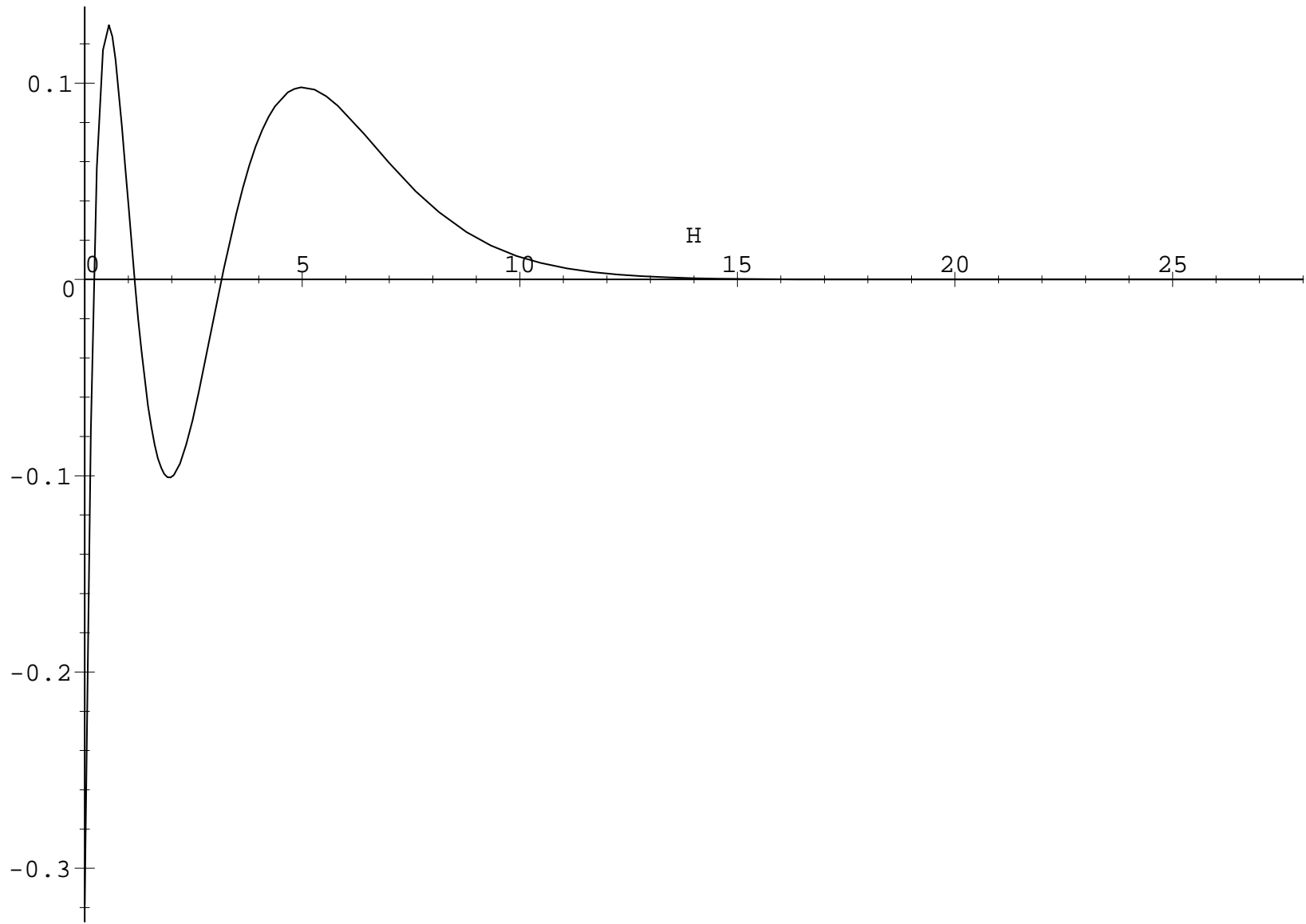
\end{document}